\documentclass[journal]{IEEEtran}
\normalsize
\ifCLASSINFOpdf
\else
\fi
\usepackage{microtype}
\DisableLigatures[f]{encoding = *, family = * }
\usepackage{array}
\usepackage{color}
\usepackage{amsfonts}
\usepackage{amssymb}
\usepackage{amsmath}
\usepackage[dvips]{graphicx}
\usepackage{epstopdf}
\usepackage{cite}
\usepackage{balance}
\usepackage{textcomp}
\usepackage{gensymb}
\usepackage{float}
\usepackage{tabularx}
\usepackage{multirow}
\usepackage{amsthm}
\usepackage{flushend}
\usepackage[caption=false]{subfig}
\usepackage{nccmath}

\newtheorem{lemma}{\it Lemma}

\makeatletter
  \def\vhrulefill#1{\leavevmode\leaders\hrule\@height#1\hfill \kern\z@}
\makeatother

\begin{document}
\title{\huge The $\alpha$-$\eta$-$\mathcal{F}$ and $\alpha$-$\kappa$-$\mathcal{F}$ Composite Fading Distributions}
\author{{Osamah. S. Badarneh}
\thanks{O. S. Badarneh is with the Electrical and Communication Engineering Department, German-Jordanian University, Amman 11180, Jordan (e-mail: Osamah.Badarneh@gju.edu.jo).}
}
\maketitle
\begin{abstract}
In this paper, we present the $\alpha$-$\eta$-$\mathcal{F}$ and $\alpha$-$\kappa$-$\mathcal{F}$ composite fading distributions. The two distributions generalize the two well-known composite fading distributions, namely the $\eta$-$\mu$/inverse gamma and the $\kappa$-$\mu$/inverse gamma distributions. For both distributions, we derive new exact closed-form expressions for the probability density function (PDF) and cumulative density function (CDF) of the instantaneous signal-to-noise ratio (SNR). Additionally, the outage probabilities (OPs) of a wireless communication system operating over the new composite fading channels are obtained. Moreover, we derive asymptotic OPs, in the high SNR regime, to gain more insight into the influence of fading parameters. Some numerical results on the envelope PDFs and OPs are provided and compared with Monte-Carlo simulation results to validate the analyses.

\end{abstract}
\begin{IEEEkeywords}
Channel modeling, composite fading, inverse gamma (Nakagami).
\end{IEEEkeywords}
\IEEEpeerreviewmaketitle
\section{Introduction}
\IEEEPARstart{W}{ireless} fading channels can strongly affect the communication between the transmitter and receiver. Hence, characterization of wireless fading channels plays a crucial role in wireless communication systems design.

Over the past decades, there is considerable research on channel fading characterization. Recently, the authors in \cite{8166770} extended the work in \cite{m:yac} and proposed two composite fading channels, namely the $\eta$-$\mu$/inverse gamma and $\kappa$-$\mu$/inverse gamma. In \cite{7460203}, the authors presented a unified model for the composite $\eta$-$\mu$/gamma, $\kappa$-$\mu$/gamma, and $\alpha$-$\mu$/gamma distributions based on a mixture gamma distribution. Based on the $\kappa$-$\mu$ model \cite{m:yac}, the authors in \cite{Paris} presented the $\kappa$-$\mu$ shadowed fading distribution. The authors in \cite{1905-00069} explored the utility of inverse gamma distribution in characterizing composite fading channels. The $\alpha$-$\kappa$-$\mu$ shadowed fading distribution was proposed as a generalization of the $\kappa$-$\mu$ shadowed fading distribution \cite{1904-05587}. In \cite{Fisher}, the authors modeled the multi-path components by Nakagami distribution, while the shadowing was modeled by inverse Nakagami distribution. The resulting model is known as Fisher-Snedecor $\mathcal{F}$ distribution. The authors in \cite{8920091} employed the gamma and the inverse gamma distributions to model the small- and large-scale irradiance variations of the propagating wave, respectively, where the resulting model is equivalent to the model proposed in \cite{Fisher}.

In this letter, two new composite fading distributions, namely the $\alpha$-$\eta$-$\mathcal{F}$ and the $\alpha$-$\kappa$-$\mathcal{F}$ distributions are presented. Unlike the existing distributions in the literature, the composite distributions proposed in this letter jointly consider the effect of shadowing and the non-linearity of the propagation medium. Moreover, the new composite distributions can offer more flexibility as they include other well-known distributions such as the $\kappa$-$\mu$/inverse gamma and the $\eta$-$\mu$/inverse gamma distributions \cite{8166770}, the $\alpha$-$\eta$-$\mu$ and the $\alpha$-$\kappa$-$\mu$ distributions \cite{4100514} --and their inclusive ones-- as special cases. Some potential applications of the proposed distributions include device-to-device, wearable, cellular and vehicular communications. For the proposed composite distributions, the exact analytical expressions for the probability density function (PDF) and cumulative distribution function (CDF) are derived. Some numerical results accompanied by Monte-Carlo simulations are provided to validate the analysis.

The remainder of this paper is organized as follows. Section \ref{sec2} describes the physical models of the proposed composite fading distributions and derives the PDFs and CDFs of the instantaneous signal-to-noise ratio (SNR) for both distributions. In Section \ref{sec3}, we provide some plots on the envelope PDFs and the outage probabilities (OPs). Finally, Section \ref{sec4} concludes this letter.

\section{The $\alpha$-$\eta$-$\mathcal{F}$ and $\alpha$-$\kappa$-$\mathcal{F}$ Composite Fading Distributions}\label{sec2}
\subsection{The $\alpha$-$\eta$-$\mathcal{F}$ Distribution}
In the $\alpha$-$\eta$-$\mathcal{F}$ composite distribution, the received signal envelope is defined as
\begin{align}\label{aef}
&R_{\alpha\eta\mathcal{F}}^{\alpha} =  Z^{2}X_{I}^{2}+Z^{2}Y_{Q}^{2} =  \sum_{i=1}^{\mu} Z^{2} X_{i}^{2}+ \sum_{i=1}^{{\mu}}Z^{2}Y_{i}^{2},&
\end{align}
where $\alpha>0$ represents the nonlinearity of the propagation media, $\mu$ represents the number of multi-path clusters, and $X_{I}$ and $Y_{Q}$ the in-phase and quadrature components, respectively. Specifically, $X_{i}$ and $Y_{i}$ are mutually independent Gaussian random variables (RVs), which represent the in-phase and quadrature components of the cluster $i$, respectively, with $\mathbb{E}[X_{i}]=\mathbb{E}[Y_{i}]=0$. Similar to the $\eta$-$\mu$ distribution \cite{m:yac},  the $\alpha$-$\eta$-$\mathcal{F}$ distribution has two formats. In Format I, $\mathbb{E}[X_{i}^{2}]=\sigma_{X}^{2}$, $\mathbb{E}[Y_{i}^{2}]=\sigma_{Y}^{2}$, and  $\eta=\sigma_{X}^{2}/\sigma_{Y}^{2}$, ($0<\eta<\infty$), represents the scattered-wave power ratio between $X_{I}$ and $Y_{Q}$ of each cluster of multi-path, whereas in Format II, $\mathbb{E}[X_{i}^{2}]=\mathbb{E}[Y_{i}^{2}]=\sigma^{2}$, and $\eta=\mathbb{E}[X_{i}Y_{i}]/\sigma^{2}$, ($-1<\eta<1$), represents the correlation coefficient between $X_{I}$ and $Y_{Q}$. Note that the two formats can be related to each other by $\eta_{\text{Format I}}=({1-\eta_{\text{Format II}})/(1+\eta_{\text{Format II}}})$, or equivalently, by $\eta_{\text{Format II}}=({1-\eta_{\text{Format I}})/(1+\eta_{\text{Format I}}})$. In \eqref{aef}, $Z$ is a normalized inverse Nakagami RV, i.e., $\mathbb{E}[Z^{2}]=1$, with shape parameter $m_s$ whose PDF is given by
\begin{align}\label{invnak}
&f_{Z}(z ) = \frac {{2{\left(m_{s}-1\right)}^{m_{s}}}}{{\Gamma \left ({ m_{s} }\right ) {z ^{2m_{s} + 1}}}} \exp \left ({ - \frac {{m_{s}}-1}{z ^{2}} }\right ),~~ m_{s}>1&
\end{align}
where $\Gamma(\cdot)$ is the gamma function \cite[Eq. (8.310.1)]{i:ryz}.


\begin{lemma}[The $\alpha$-$\eta$-$\mathcal{F}$ distribution: PDF]
Let $\gamma\in\mathbb{R}^{+}$ be a RV that represents the instantaneous SNR under $\alpha$-$\eta$-$\mathcal{F}$ distribution, with $\overline{\gamma}=\mathbb{E}[{\gamma}]\in\mathbb{R}^{+}$ and shape parameters $\alpha, \eta, \mu \in\mathbb{R}^{+}$ and  $m_{s}>1$, i.e., $\gamma\sim\mathcal{G_{\alpha\eta\mathcal{F}}}\left(\overline{\gamma},\alpha, \eta,\mu, m_{s}\right)$. Therefore, the PDF of the instantaneous SNR $\gamma$ can be expressed as in \eqref{pdfgam}, at the top of the next page.
\begin{fleqn}
\begin{figure*}
\begin{align}\label{pdfgam}
&f_{\gamma_{\alpha\eta\mathcal{F}}}(\gamma) = {\alpha2^{2\mu-1}\mu^{2\mu}h^{\mu}\left((m_s-1)\upsilon\overline{\gamma}^{\alpha\over2}\right)^{m_s}\gamma^{\alpha\mu-1}\over B\left(2\mu,m_s\right)\left(2\mu h \gamma^{\alpha\over2} +(m_s-1)\upsilon\overline{\gamma}^{\alpha\over2} \right)^{2\mu+m_s}}{}_{2}F_{1}\Bigg(\mu+{m_s\over2},\mu +{m_s+1\over2};\mu+{1\over2};\frac{\left(2\mu H \gamma^{\alpha\over2}\right)^{2}}{\big(2\mu h \gamma^{\alpha\over2} +(m_s-1)\upsilon\overline{\gamma}^{\alpha\over2} \big)^{2}}\Bigg),&
\cr&
\text{where}\qquad
\upsilon = \left({2\mu h\over m_s-1}\right)\Bigg({{B}\left(2\mu,m_s\right)h^{\mu}\over {B}\left(2\mu+{2\over\alpha},m_s-{2\over\alpha}\right){}_{2}F_{1}\left(\mu+{1\over\alpha},\mu +{1\over\alpha}+{1\over2};\mu+{1\over2};\frac{H^{2}}{h^{2}}\right)}\Bigg)^{\alpha\over2}.
\end{align}
\hrulefill
\end{figure*}
\end{fleqn}

\rm{where in \eqref{pdfgam}, $B(\cdot,\cdot)$ is the beta function \cite{i:ryz} and ${}_{2}F_{1}(\cdot)$ is the Gauss hypergeometric function \cite[Eq. (9.111)]{i:ryz}, $h=(2+\eta^{-1} + \eta)/4$ and $H=(\eta^{-1} - \eta)/4$ in Format I, whereas $h={1/(1-\eta^{-2})}$ and $H={\eta/(1-\eta^{-2})}$ in Format II.}
\begin{figure*}
\begin{align}\label{cpdfgam}
&F_{\gamma_{\alpha\eta\mathcal{F}}}(\gamma)  = {2^{2\mu-1}h^{\mu}\over \Gamma(2\mu)\Gamma(m_s)}\sum_{k=0}^{\infty}{\Gamma(2\mu+2k+m_s)H^{2k}\over k!{\left(\mu+{1\over2}\right)}_{k}(\mu+k)}\cr&\qquad\qquad\qquad\qquad\qquad\qquad\times\left({\mu\gamma^{\alpha\over2}\over(m_s-1)\upsilon\overline{\gamma}^{\alpha\over2}}\right)^{2\mu+2k}
{}_{2}F_{1}\Bigg(2\mu+2k+m_s,2\mu +2k;2\mu+2k+1;\frac{-2\mu h \gamma^{\alpha\over2}}{(m_s-1)\upsilon\overline{\gamma}^{\alpha\over2} }\Bigg).&
\end{align}
\hrulefill
\end{figure*}
\end{lemma}
\begin{IEEEproof}
See Appendix A.1.
\end{IEEEproof}
\begin{figure*}
\begin{align}\label{cpdftrun3}\begin{aligned}
&T_{\epsilon}  \leq{}_{2}F_{1}\Bigg(2\mu+2K_{0}+m_s,2\mu +2K_{0};2\mu+2K_{0}+1;\frac{-2\mu h \gamma^{\alpha\over2}}{(m_s-1)\upsilon\overline{\gamma}^{\alpha\over2} }\Bigg)\left({\mu\gamma^{\alpha\over2}\over(m_s-1)\upsilon\overline{\gamma}^{\alpha\over2}}\right)^{2\mu}\cr&\qquad\qquad\qquad\qquad\qquad\qquad\qquad\qquad\times
{\Gamma(m_{s}+2\mu)\over\left(\mu+K_{0}\right)}{}_{2}F_{1}\Bigg({2\mu+m_{s}\over2},{2\mu +m_{s}+1\over2};{2\mu+1\over2};\left(\frac{2\mu H \gamma^{\alpha\over2}}{(m_s-1)\upsilon\overline{\gamma}^{\alpha\over2} }\right)^{2}\Bigg).&
\end{aligned}\end{align}
\hrulefill
\end{figure*}
\begin{lemma}[The $\alpha$-$\eta$-$\mathcal{F}$ distribution: CDF]
The CDF of the instantaneous SNR $\gamma$ under $\alpha$-$\eta$-$\mathcal{F}$ distribution can be derived as in \eqref{cpdfgam}, where $\left(x\right)_{n}$ denotes the Pochhammer symbol \cite{i:ryz}.
\end{lemma}
\begin{IEEEproof}
See Appendix A.2.
\end{IEEEproof}

\begin{lemma}[Truncation error for the CDF in \eqref{cpdfgam}: $T_{\epsilon}$] The truncation error for the CDF in \eqref{cpdfgam}, if it is truncated after $K_{0}-1$ terms, can be given in closed-form as in \eqref{cpdftrun3}.
\end{lemma}
\begin{IEEEproof}
See Appendix A.3.
\end{IEEEproof}

Note that when $m_{s}\rightarrow\infty$, the $\alpha$-$\eta$-$\mathcal{F}$ distribution coincides with the $\alpha$-$\eta$-$\mu$ distribution. Moreover, when $\alpha=2$, the $\alpha$-$\eta$-$\mathcal{F}$ distribution is equivalent to the $\eta$-$\mu$/inverse gamma distribution and hence; all special cases in \cite[Table I]{8166770} can be deduced. Additionally, when $\eta\rightarrow1$, the $\alpha$-$\eta$-$\mathcal{F}$ reduces to $\alpha$-$\mathcal{F}$ distribution. Through the $\alpha$-$\mathcal{F}$ distribution, the Fisher-Snedecor $\mathcal{F}$ distribution can be deduced when $\alpha=2$. When $\mu=1$, the $\alpha$-$\eta$-$\mathcal{F}$ reduces to $\alpha$-$\eta$/inverse gamma. To the best of authors' knowledge, the $\alpha$-$\mathcal{F}$ and the $\alpha$-$\eta$/inverse gamma distributions have been not reported in the literature before.
\subsection{The $\alpha$-$\kappa$-$\mathcal{F}$ Distribution}
In the $\alpha$-$\kappa$-$\mathcal{F}$ composite distribution, the received signal envelope is defined as
\begin{align}\label{akf}
&R_{\alpha\kappa\mathcal{F}}^{\alpha} =  \sum_{i=1}^{\mu} Z^{2} (X_{i}+p_{i})^{2}+ \sum_{i=1}^{\mu}Z^{2}(Y_{i}+q_{i})^{2},&
\end{align}
where $\alpha$, $\mu$, $Z$, $X_{i}$, and $Y_{i}$ are defined as in \eqref{aef}, while here $\mathbb{E}[X_{i}^{2}]=\mathbb{E}[Y_{i}^{2}]=\sigma^{2}$. $p_{i}$ and $q_{i}$ are defined as in \cite{m:yac}, that is, $p_{i}$ and $q_{i}$ represent the mean values of the in-phase and quadrature components of the multipath waves of cluster $i$, respectively. Note that the ratio of the total power of the dominant components to the
total power of the scattered waves represents the parameter $\kappa$. As such, $\kappa=d^{2}/2\mu\sigma^{2}>0$, where $d^{2}=\sum_{i=1}^{\mu}p_{i}^{2}+q_{i}^{2}$.

\begin{lemma}[The $\alpha$-$\kappa$-$\mathcal{F}$ distribution: PDF]
Let $\gamma\in\mathbb{R}^{+}$ be a RV that represents the instantaneous SNR under $\alpha$-$\kappa$-$\mathcal{F}$ distribution, with $\overline{\gamma}=\mathbb{E}[{\gamma}]\in\mathbb{R}^{+}$ and shape parameters $\alpha, \kappa, \mu \in\mathbb{R}^{+}$ and $m_{s}>1$, i.e., $\gamma\sim\mathcal{G_{\alpha\kappa\mathcal{F}}}\left(\overline{\gamma},\alpha, \kappa,\mu, m_{s}\right)$. Therefore, the PDF of the instantaneous SNR $\gamma$ can be written as
\begin{align}\label{pdfgamk}
&f_{\gamma_{\alpha\kappa\mathcal{F}}}(\gamma) = {\alpha\mu^{\mu}(1+\kappa)^{\mu}\left((m_s-1)\omega\overline{\gamma}^{\alpha\over2}\right)^{m_s}\exp{(-\mu\kappa)}\over 2{B}\left(\mu,m_s\right)\left(\mu (1+\kappa) \gamma^{{\alpha\over2}} +(m_s-1)\omega\overline{\gamma}^{\alpha\over2} \right)^{\mu+m_s}}
\cr&\times \gamma^{{\alpha\mu\over2}-1}{}_{1}F_{1}\left(\mu+m_s;\mu;\frac{\mu^{2} \kappa(1+\kappa) \gamma^{{\alpha\over2}}}{\mu(1+\kappa)\gamma^{{\alpha\over2}} +(m_s-1)\omega\overline{\gamma}^{\alpha\over2}}\right),&
\end{align}
where ${}_{1}F_{1}(\cdot)$ is the Confluent hypergeometric function \cite[Eq. (9.210.1)]{i:ryz} an $\omega$ is defined as
\begin{align}\label{ups}
&\omega = {\mu (1+\kappa)\over m_s-1}\Bigg({{B}\left(\mu,m_s\right)\exp{(\mu\kappa)}\over {B}\left(\mu+{2\over\alpha},m_s-{2\over\alpha}\right){}_{1}F_{1}\left(\mu+{2\over\alpha};\mu ;\mu\kappa\right)}\Bigg)^{\alpha\over2}.&
\end{align}
\end{lemma}
\begin{IEEEproof}
See Appendix A.3.
\end{IEEEproof}

\begin{lemma}[The $\alpha$-$\kappa$-$\mathcal{F}$ distribution: CDF]
The CDF of the instantaneous SNR $\gamma$ under $\alpha$-$\kappa$-$\mathcal{F}$ distribution can be derived as
\begin{align}\label{cpdfgamk}\begin{aligned}
&F_{\gamma_{\alpha\kappa\mathcal{F}}}(\gamma)  =  \sum_{t=0}^{\infty}{\exp{(-\mu\kappa)}(\mu\kappa)^{t}\over t!\,{\rm B}(\mu+t,m_{s})(\mu+t)}\left({\mu(1+\kappa)\gamma^{\alpha\over2}\over(m_s-1)\omega\overline{\gamma}^{\alpha\over2}}\right)^{\mu+t}
\cr&\times{}_{2}F_{1}\Bigg(\mu+m_s+t,\mu +t;\mu+t+1;{-\mu(1+\kappa)\gamma^{\alpha\over2}\over(m_s-1)\omega\overline{\gamma}^{\alpha\over2}}\Bigg).&
\end{aligned}\end{align}
For the case $(m_s-1)\omega\overline{\gamma}^{\alpha\over2}>\mu(1+\kappa)\gamma^{\alpha\over2}$, we can rewrite \eqref{cpdfgamk} in closed-form, i.e., in terms of Kamp\'{e} de F\'{e}riet function ${F^{{p;q;r}}_{k;m;n}}[\cdot]$ \cite[Eq. (1.3.28)]{Srivastava}, as
\begin{align}\label{kamp}\begin{aligned}
&F_{\gamma_{\alpha\kappa\mathcal{F}}}(\gamma)={\exp{(-\mu\kappa)}\over{\mu\,\rm B}(\mu,m_{s})}\left({\mu(1+\kappa)\gamma^{\alpha\over2}\over(m_s-1)\omega\overline{\gamma}^{\alpha\over2}}\right)^{\mu}\cr&\times
F \mathstrut_{1:1;0}^{2:0;0}
    \left[
      \begin{array}{@{} c @{:} c @{;} c @{;}}
        {\mu + m_{s}, \mu} &
          \frac{\phantom{-}}{\phantom{-}} &
           \frac{\phantom{-}}{\phantom{-}} \\
        {\mu + 1} &
          {\mu} &
          \frac{\phantom{-}}{\phantom{-}}
      \end{array}
      {\mu^{2}\kappa(1+\kappa)\gamma^{\alpha\over2}\over(m_s-1)\omega\overline{\gamma}^{\alpha\over2}},-{\mu(1+\kappa)\gamma^{\alpha\over2}\over(m_s-1)\omega\overline{\gamma}^{\alpha\over2}}
    \right],&
  \end{aligned}\end{align}
while for the case $(m_s-1)\omega\overline{\gamma}^{\alpha\over2}\leq\mu(1+\kappa)\gamma^{\alpha\over2}$, \eqref{cpdfgamk} can be rewritten in terms of Humbert $\Psi_{1}$ function \cite[Eq. (1.3.21)]{Srivastava} as
\begin{align}\label{hum}\begin{aligned}
&F_{\gamma_{\alpha\kappa\mathcal{F}}}(\gamma)=e^{-\mu\kappa}\Psi_{1}\Bigg(\mu;0;1-m_{s},\mu,{-(m_s-1)\omega\overline{\gamma}^{\alpha\over2}\over\mu(1+\kappa)\gamma^{\alpha\over2}},\mu\kappa\Bigg)\cr&-
\Bigg[ {\exp{(-\mu\kappa)}\over{m_{s}\,\rm B}(\mu,m_{s})}\left((m_s-1)\omega\overline{\gamma}^{\alpha\over2}\over\mu(1+\kappa)\gamma^{\alpha\over2}\right)^{m_{s}}
\cr&\qquad\times\Psi_{1}\Bigg(\mu+m_{s};m_{s};1+m_{s},\mu,{-(m_s-1)\omega\overline{\gamma}^{\alpha\over2}\over\mu(1+\kappa)\gamma^{\alpha\over2}},\mu\kappa\Bigg)\Bigg].&
\end{aligned}\end{align}

\end{lemma}
\begin{IEEEproof}
See Appendix A.5.
\end{IEEEproof}
Note that when $m_{s}\rightarrow\infty$, the $\alpha$-$\kappa$-$\mathcal{F}$ distribution coincides with the $\alpha$-$\kappa$-$\mu$ distribution. Moreover, when $\alpha=2$, the $\alpha$-$\kappa$-$\mathcal{F}$ distribution is equivalent to the $\kappa$-$\mu$/inverse gamma distribution and hence; all special cases in \cite[Table I]{8166770} can be deduced. Additionally, when $\kappa\rightarrow0$, the $\alpha$-$\kappa$-$\mathcal{F}$ reduces to $\alpha$-$\mathcal{F}$ distribution, whereas when $\mu=1$, the $\alpha$-$\kappa$-$\mathcal{F}$ reduces to $\alpha$-$\kappa$/inverse gamma. To the best of authors' knowledge, the $\alpha$-$\kappa$/inverse gamma distribution has been not reported in the literature before.
\section{Simulation Results and Discussion}\label{sec3}
To validate our analysis, we provide in this section some plots for the envelope PDFs of the both distributions. Also, we provide some results on the outage probability (OP) to validate the derived CDFs and to study the impact of different channels parameters.

\begin{figure}[t!] \centering
\hspace{-0.2in}
         \includegraphics[width=18pc,keepaspectratio]{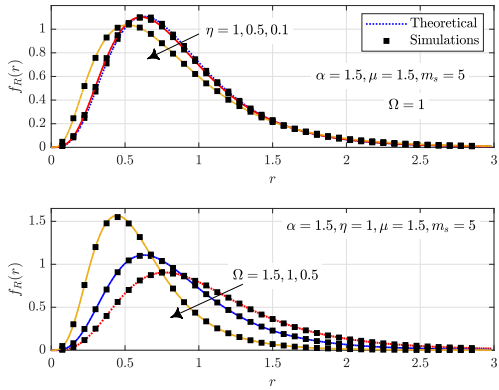}%
       \caption{The $\alpha$-$\eta$-$\mathcal{F}$ PDF distribution for different values of $\mu$ and $\Omega$.}\label{pdfeid}
\end{figure}

\begin{figure}[t!] \centering
\hspace{-0.2in}
         \includegraphics[width=18pc,keepaspectratio]{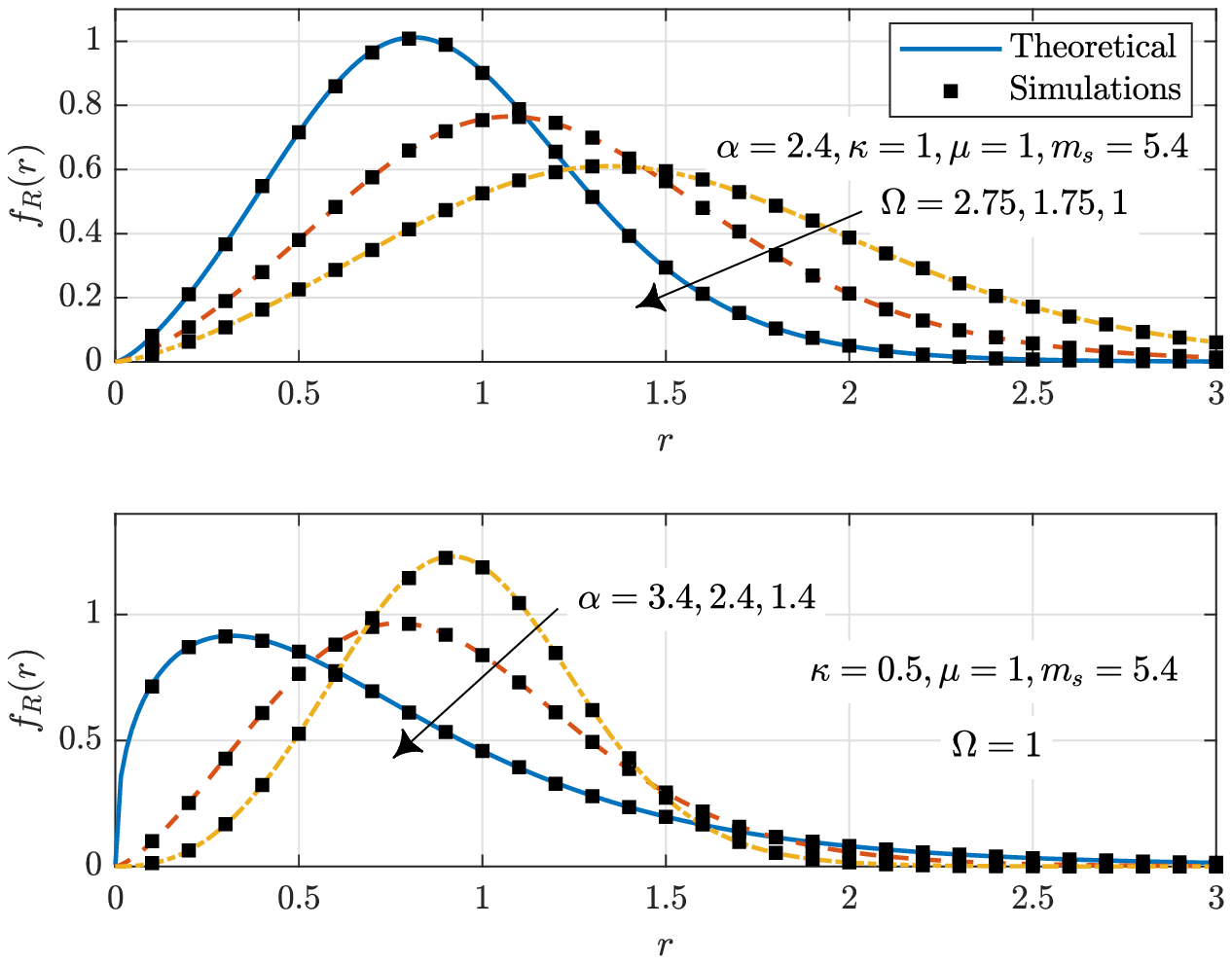}%
       \caption{The $\alpha$-$\kappa$-$\mathcal{F}$ PDF distribution for different values of $\kappa$ and $\alpha$.}\label{pdfemp}
\end{figure}

\begin{figure}[t!] \centering
\hspace{-0.2in}
         \includegraphics[width=18pc,keepaspectratio]{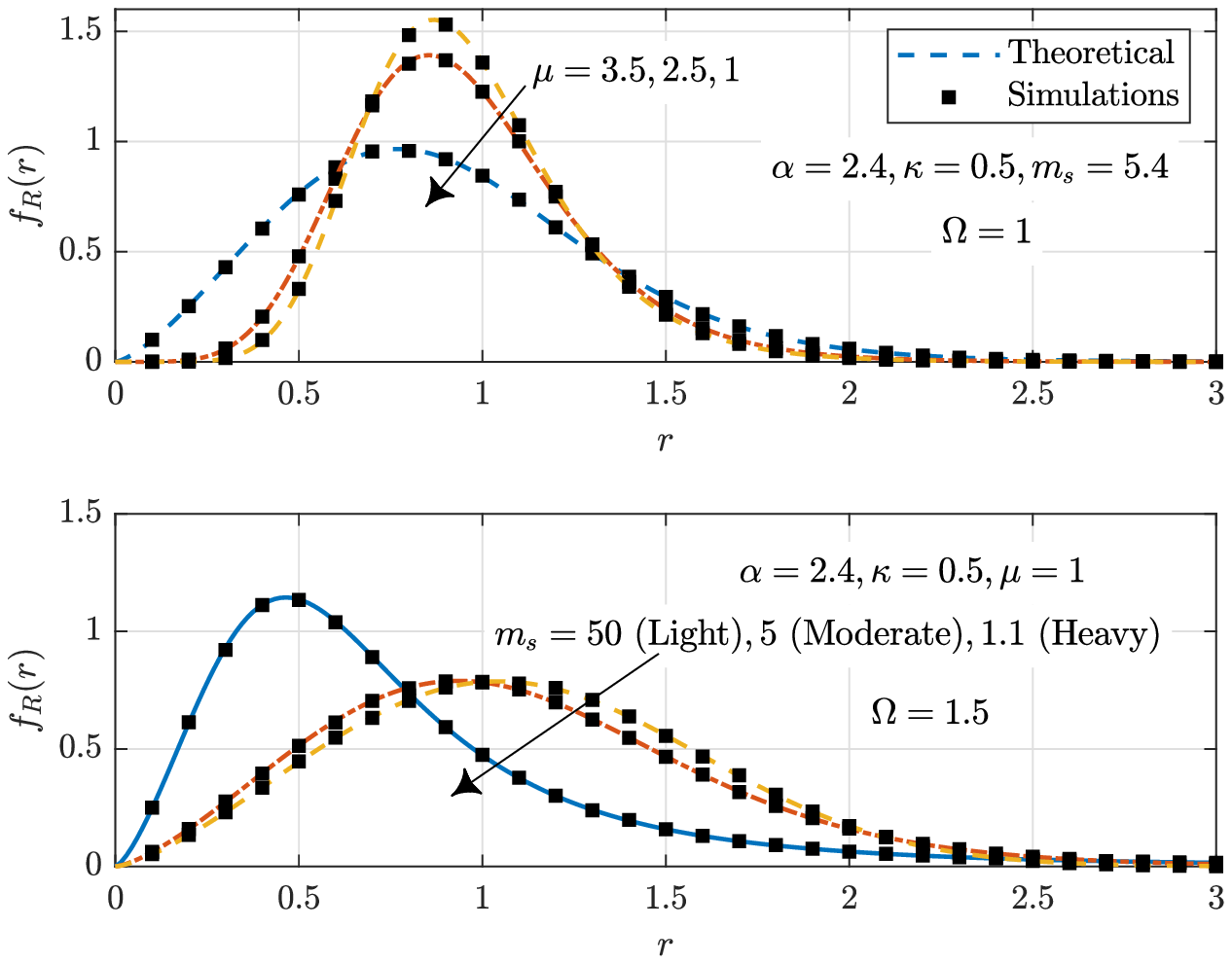}%
       \caption{The $\alpha$-$\kappa$-$\mathcal{F}$ PDF distribution for different values of $\mu$ and $m_s$.}\label{pdfkid}
\end{figure}

Figs. \ref{pdfeid}-\ref{pdfkid} show the envelope PDFs of the $\alpha$-$\eta$-$\mathcal{F}$, Format I, and the $\alpha$-$\kappa$-$\mathcal{F}$ distributions for different parameters. Clearly, there is an excellent agreement between the numerical results and the Monte-Carlo simulations, which validates the derived PDFs. Specifically, Fig. \ref{pdfeid} shows that the PDF curves are less sensitive with respect to the value of $\eta$, which means that increasing $\eta$ has less effect on the system performance. Additionally, Fig. \ref{pdfeid} and Fig. \ref{pdfemp} show that lower values of $\Omega$ shift the curves closer to the ordinate axis. Moreover, Fig. \ref{pdfemp} and Fig. \ref{pdfkid} show that higher values of the parameters $\alpha$, $\mu$ or $m_{s}$ tend to produce PDF curves more concentrated around the mean $\Omega$, which means better performance.

\begin{figure}[!t] \centering
\hspace{-0.2in}
         \includegraphics[width=18.5pc,keepaspectratio]{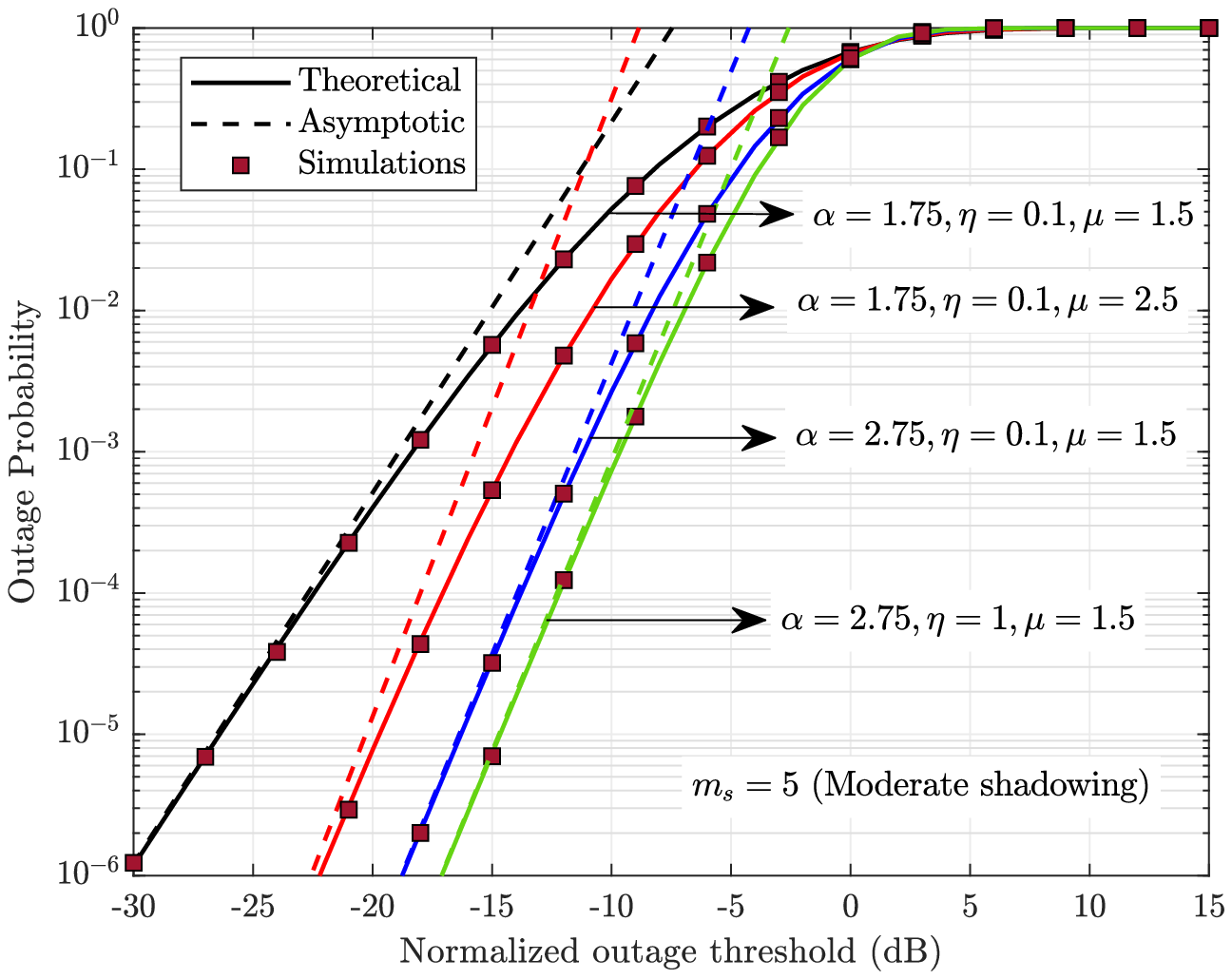}%
       \caption{OP as a function of normalized outage threshold over $\alpha$-$\eta$-$\mathcal{F}$ model.}\label{outaef}
\end{figure}
\begin{figure}[!t] \centering
\hspace{-0.2in}
         \includegraphics[width=16.0pc,keepaspectratio]{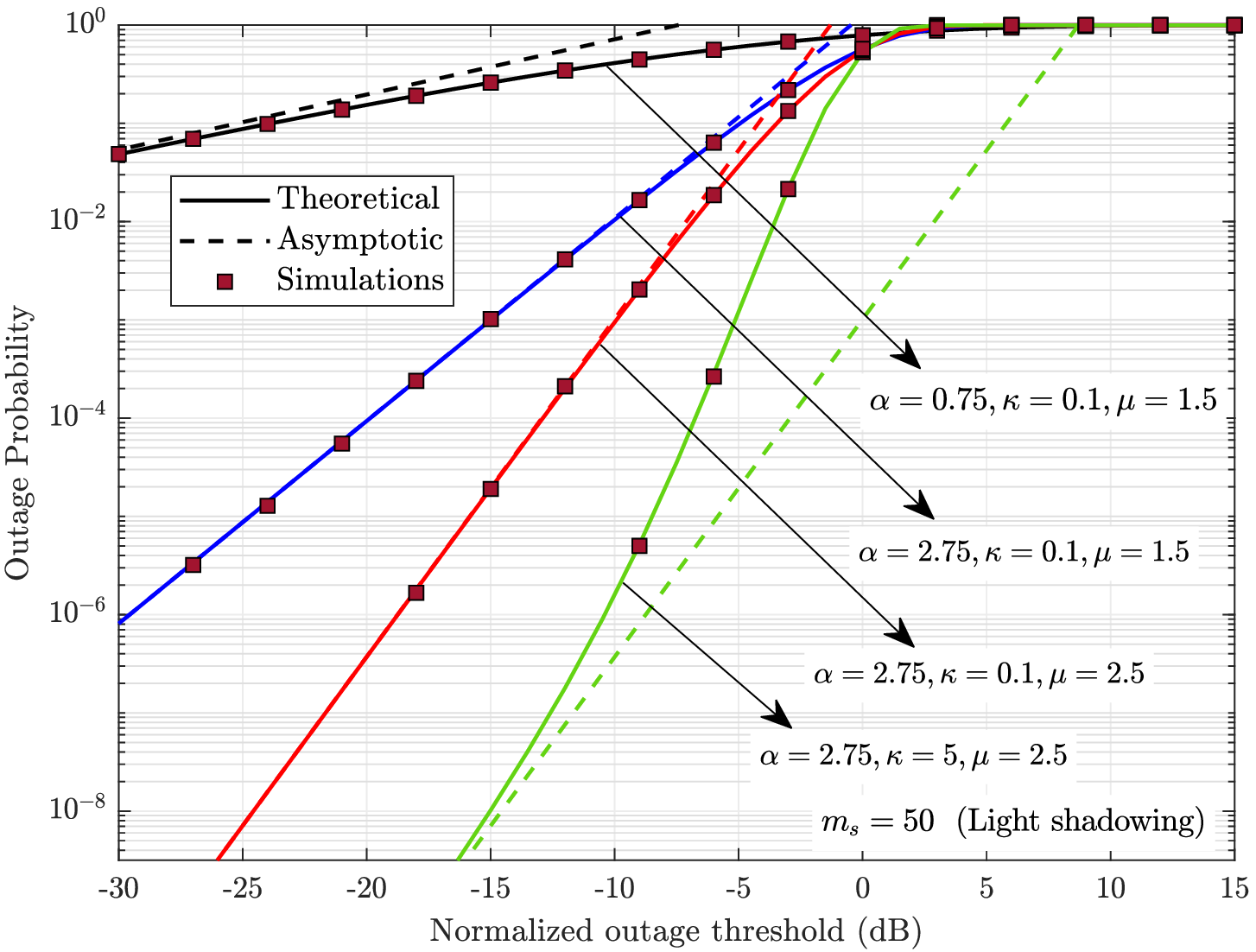}%
       \caption{OP as a function of normalized outage threshold over $\alpha$-$\kappa$-$\mathcal{F}$ model.}\label{outakm}
\end{figure}

The OP can be evaluated using $\mathcal{OP}_{\Xi}=F_{\Xi}(\gamma_{\rm th})$, where $\Xi\in\{\alpha\eta\mathcal{F},\alpha\kappa\mathcal{F}\}$, $F_{\Xi}(\cdot)$ is given in \eqref{cpdfgam} and \eqref{cpdfgamk}. The asymptotic OP can be obtained when $\overline{\gamma}\rightarrow\infty$ and using the fact that ${_{2}F_{1}} \left (\cdot,\cdot;\cdot;0\right)=1$, thus we get
\begin{align}\label{cpdfgamasy}
&\mathcal{OP}_{\gamma_{\alpha\eta\mathcal{F}}}= {(2\mu)^{2\mu-1}h^{\mu}\over {\rm{B}}(2\mu,m_s)}\left({\gamma_{\rm th}^{\alpha\over2}\over(m_s-1)\upsilon\overline{\gamma}^{\alpha\over2}}\right)^{2\mu}.&
\end{align}
In the high SNR regime, the OP can be expressed as $P_{\rm{out}}=(G_{c}\overline{\gamma})^{-G_{d}}+\mathcal{O}(\overline{\gamma}^{-G_{d}})$ where $G_{c}$ and $G_{d}$ are the coding and diversity gains. It is obvious from \eqref{cpdfgamasy} that $G_{c}={1\over\gamma_{\rm th}}\left({(2\mu)^{2\mu-1}h^{\mu}\over{\rm{B}}(2\mu,m_s)(m_s -1)^{2\mu}\upsilon^{2\mu}}\right)^{-{1\over\alpha\mu}}$ and $G_{d}={\alpha\mu}$.

For the $\alpha$-$\kappa$-$\mathcal{F}$ distribution, the asymptotic OP can be derived as
\begin{align}\label{cpdfgamkasy}
&\mathcal{OP}_{\alpha\kappa\mathcal{F}}= {\mu^{\mu-1}\exp{(-\mu\kappa)}\over{\rm B}(\mu,m_{s})}\left({(1+\kappa)\gamma_{\rm th}^{\alpha\over2}\over(m_s-1)\omega\overline{\gamma}^{\alpha\over2}}\right)^{\mu}
.&
\end{align}
It is clear that $G_c={1\over\gamma_{\rm th}}\left({\mu^{\mu-1}\exp{(-\mu\kappa)}(1+\kappa)^{\mu}\over{\rm B}(\mu,m_{s})(m_s-1)^{\mu}}\right)^{-{2\over\alpha\mu}}$ and $G_d={\alpha\mu\over2}$.
The OP, for both channels, as a function of normalized outage threshold $\gamma_{\rm th}/\overline{\gamma}$ is depicted in Fig. \ref{outaef} and Fig. \ref{outakm}. It is clear that as any of the parameters $\alpha$, $\eta$, $\kappa$, and $\mu$ increases, the OP performance improves. However, the impact of $\alpha$ and $\mu$ is more pronounced of that of $\eta$ and $\kappa$; which can be related to the different phenomena that each parameter describes. Additionally, Fig. \ref{outaef} and Fig. \ref{outakm} show that the slope of the asymptotic curves increases as $\alpha$ or $\mu$ increases. On the other hand, the slope does not change as $\eta$ or $\kappa$ changes; which confirms the diversity gain in \eqref{cpdfgamasy} and \eqref{cpdfgamkasy}.

\section{Conclusions}\label{sec4}
Two composite fading distributions were presented in this letter, namely the  $\alpha$-$\eta$-$\mathcal{F}$ and $\alpha$-$\kappa$-$\mathcal{F}$ distributions. The  $\alpha$-$\eta$-$\mathcal{F}$ and $\alpha$-$\kappa$-$\mathcal{F}$ distributions are very versatile as various well-known fading channels in the literature can be obtained as special cases. Since the new distributions jointly consider the non-linearity of the propagation media and shadowing, it is expected that they can provide better flexibility to characterize wireless fading channels compared with other distributions in the literature, such as the $\alpha$-$\eta$-$\mu$, $\alpha$-$\kappa$-$\mu$, $\eta$-$\mu$/inverse gamma, and $\kappa$-$\mu$/inverse gamma distributions.
\begin{appendices}
\section{}
\subsubsection*{A.1) Proof of Lemma 1}
Let's define the received signal envelope of $\eta$-$\mathcal{F}$ composite distribution as
\begin{align}\label{ef}
  &R_{\eta\mathcal{F}}^{2} =\sum_{i=1}^{{\mu}} Z^{2} X_{i}^{2}+ \sum_{i=1}^{{\mu}}Z^{2}Y_{i}^{2}.&
  \end{align}

Therefore, the envelope PDF of the $\eta$-$\mathcal{F}$ composite distribution (equivalently, the $\eta$-$\mu$/inverse gamma \cite{8166770}), can be derived as
\begin{align}\label{pdfef}
&f_{R_{\eta\mathcal{F}}}(r) = {2^{2\mu+1}\mu^{2\mu}h^{\mu}\left((m_s-1)\lambda\right)^{m_s}\over {B}\left(2\mu,m_s\right)\left(2\mu h r^2 +(m_s-1)\lambda \right)^{2\mu+m_s}}r^{4\mu-1}\cr&{}_{2}F_{1}\Bigg(\mu+{m_s\over2},\mu +{m_s+1\over2};\mu+{1\over2};\frac{\left(2\mu H r^2\right)^{2}}{\big(2\mu h r^2 +(m_s-1)\lambda \big)^{2}}\Bigg).&
\end{align}
where $\lambda$ is the mean signal power which is given as $\lambda=\mathbb{E}[R_{\eta\mathcal{F}}^{2}]=2\mu(1+\eta^{-1})\sigma_{X}^{2}=2\mu(1+\eta)\sigma_{Y}^{2}$ in Format I and  $\lambda= 2\mu\sigma^{2}$ in Format II.

It is clear from \eqref{aef} and \eqref{ef} that $R_{\alpha\eta\mathcal{F}}^{\alpha}=R_{\eta\mathcal{F}}^{2}$ with mean signal power $\Omega=\mathbb{E}[R_{\alpha\eta\mathcal{F}}^{2}]=\mathbb{E}[R_{\eta\mathcal{F}}^{4\over\alpha}]$. Using the concept of transformation of RVs, the envelope PDF of the $\alpha$-$\eta$-$\mathcal{F}$ composite distribution can be derived as in \eqref{aef1}.
\begin{figure*}
\begin{align}\label{aef1}
&f_{R_{\alpha\eta\mathcal{F}}}(r) = {\alpha2^{2\mu}\mu^{2\mu}h^{\mu}\left((m_s-1)\upsilon\Omega^{\alpha\over2}\right)^{m_s}r^{2\alpha\mu-1}\over {B}\left(2\mu,m_s\right)\left(2\mu h r^{\alpha} +(m_s-1)\upsilon\Omega^{\alpha\over2} \right)^{2\mu+m_s}}{}_{2}F_{1}\Bigg(\mu+{m_s\over2},\mu +{m_s+1\over2};\mu+{1\over2};\frac{\left(2\mu H r^{\alpha}\right)^{2}}{\big(2\mu h r^{\alpha} +(m_s-1)\upsilon\Omega^{\alpha\over2} \big)^{2}}\Bigg).&
\end{align}
\hrulefill
\end{figure*}
The instantaneous SNR $\gamma$ under $\alpha$-$\eta$-$\mathcal{F}$ distribution can be derived using $\gamma={\overline{\gamma}R_{\alpha\eta\mathcal{F}}^{2}\over\mathbb{E}[R_{\alpha\eta\mathcal{F}}^{2}]}={\overline{\gamma}R_{\alpha\eta\mathcal{F}}^{2}\over\Omega}$. Thus, using the concept of transformation of RVs and with the help of \eqref{aef1}, the PDF in \eqref{pdfgam} is obtained. Thus, the proof is completed.

\subsubsection*{A.2) Proof of Lemma 2}
Using $F_{\gamma}(\gamma)\triangleq\int_{0}^{\gamma}f_{\gamma}(x)dx$, \cite[3.194.1]{i:ryz}, \cite[8.335.1]{i:ryz}, and after some mathematical manipulations, then \eqref{cpdfgam} is obtained, which completes the proof.
\subsubsection*{A.3) Proof of Lemma 3}
We note that \eqref{cpdfgam} is a convergent infinite series. As such, the truncation error, $T_{\epsilon}$, if it is truncated after $K_{0}-1$ terms can be expressed as
\begin{align}\label{cpdftrun}\begin{aligned}
&T_{\epsilon}  =\sum_{k=K_{0}}^{\infty}{\Gamma(2\mu+2k+m_s)H^{2k}\over k!{\left(\mu+{1\over2}\right)}_{k}(\mu+k)}\left({\mu\gamma^{\alpha\over2}\over(m_s-1)\upsilon\overline{\gamma}^{\alpha\over2}}\right)^{2\mu+2k}
\cr&{}_{2}F_{1}\Bigg(2\mu+2k+m_s,2\mu +2k;2\mu+2k+1;\frac{-2\mu h \gamma^{\alpha\over2}}{(m_s-1)\upsilon\overline{\gamma}^{\alpha\over2} }\Bigg).&
\end{aligned}\end{align}
Capitalizing on the fact that ${}_{2}F_{1}(\cdot)$ is monotonically decreasing with respect to $k$, \eqref{cpdftrun} can be upper bounded as in \eqref{cpdftrun1}, at the top of the next page.
\begin{figure*}
\begin{align}\label{cpdftrun1}\begin{aligned}
&T_{\epsilon}  \leq{}_{2}F_{1}\Bigg(2\mu+2K_{0}+m_s,2\mu +2K_{0};2\mu+2K_{0}+1;\frac{-2\mu h \gamma^{\alpha\over2}}{(m_s-1)\upsilon\overline{\gamma}^{\alpha\over2} }\Bigg){\left({\mu\gamma^{\alpha\over2}\over(m_s-1)\upsilon\overline{\gamma}^{\alpha\over2}}\right)^{2\mu}\over(\mu+K_{0})}
\cr&\qquad\qquad\qquad\qquad\qquad\qquad\qquad\qquad\qquad\qquad\qquad\qquad\qquad\times\sum_{k=K_{0}}^{\infty}{\Gamma(2\mu+2k+m_s)H^{2k}\over k!{\left(\mu+{1\over2}\right)}_{k}}\left({\mu\gamma^{\alpha\over2}\over(m_s-1)\upsilon\overline{\gamma}^{\alpha\over2}}\right)^{2k}.&
\end{aligned}\end{align}
\hrulefill
\end{figure*}

Since strictly positive terms are added up, the summation limits in \eqref{cpdftrun1} can be expressed as
\begin{align}\label{cpdftrun2}\begin{aligned}
&\sum_{k=K_{0}}^{\infty}{\Gamma(2\mu+2k+m_s)\over k!{\left(\mu+{1\over2}\right)}_{k}}\left({\mu H\gamma^{\alpha\over2}\over(m_s-1)\upsilon\overline{\gamma}^{\alpha\over2}}\right)^{2k}\cr&\leq\sum_{k=0}^{\infty}{\Gamma(2\mu+2k+m_s)\over k!{\left(\mu+{1\over2}\right)}_{k}}\left({\mu H\gamma^{\alpha\over2}\over(m_s-1)\upsilon\overline{\gamma}^{\alpha\over2}}\right)^{2k}.&
\end{aligned}\end{align}
Applying the doubling formula \cite[Eq. (8.335.1)]{i:ryz} to the term $\Gamma(ms+2k+2\mu)$, the right-side of \eqref{cpdftrun2} can be rewritten as
\begin{align}\label{cpdftrun4}\begin{aligned}
&\leq\sum_{k=0}^{\infty}{2^{2\mu+2k+m_{s}-1}\Gamma(\mu+k+{m_s\over2})\Gamma(\mu+k+{m_s\over2}+{1\over2})\over \sqrt{\pi}k!{\left(\mu+{1\over2}\right)}_{k}}\cr&\qquad\qquad\qquad\qquad\qquad\qquad\qquad\times\left({2\mu H\gamma^{\alpha\over2}\over(m_s-1)\upsilon\overline{\gamma}^{\alpha\over2}}\right)^{2k},&
\end{aligned}\end{align}
which can be further rewritten as
\begin{fleqn}
\begin{align}\label{cpdftrun5}
&\Gamma(m_{s}+2\mu)\leq\sum_{k=0}^{\infty}{\left({2\mu+m_s\over2}\right)_{k}\left({2\mu+m_s+1\over2}\right)_{k}\left({2\mu H\gamma^{\alpha\over2}\over(m_s-1)\upsilon\overline{\gamma}^{\alpha\over2}}\right)^{2k}\over k!{\left(\mu+{1\over2}\right)}_{k}}.&
\end{align}
\end{fleqn}
Using \cite[Eq. (9.14.1)]{i:ryz}, \eqref{cpdftrun5} can be expressed in closed-form in terms of ${}_{2}F_{1}(\cdot)$. Using this result in \eqref{cpdftrun1}, then the truncation error $T_{\epsilon}$ is obtained in closed-form as in \eqref{cpdftrun3}. Note that
when $(m_s-1)\upsilon\overline{\gamma}^{\alpha\over2}<2\mu H\gamma^{\alpha\over2}$, the transformation formula can be used \cite[Eq. (9.131.1)]{i:ryz}.

\subsubsection*{A.4) Proof of Lemma 4}
Let's define the received signal envelope of $\kappa$-$\mathcal{F}$ composite distribution as
\begin{align}\label{kf}
&R_{\kappa\mathcal{F}}^{2} =  \sum_{i=1}^{\mu} Z^{2} (X_{i}+p_{i})^{2}+ \sum_{i=1}^{\mu}Z^{2}(Y_{i}+q_{i})^{2}.&
  \end{align}

Therefore, the envelope PDF of the $\kappa$-$\mathcal{F}$ composite distribution (equivalently, the $\kappa$-$\mu$/inverse gamma \cite{8166770}), can be derived as
\begin{align}\label{pdfkf}
&f_{R_{\kappa\mathcal{F}}}(r) = {2\mu^{\mu}(1+\kappa)^{\mu}\left((m_s-1)\mho\right)^{m_s}\exp{(-\mu\kappa)}\over {B}\left(\mu,m_s\right)\left(\mu (1+\kappa) r^2 +(m_s-1)\mho \right)^{\mu+m_s}}r^{2\mu-1}\cr&\qquad\qquad\times{}_{1}F_{1}\left(\mu+m_s;\mu;\frac{\mu^{2} \kappa(1+\kappa) r^2}{\mu(1+\kappa)r^2 +(m_s-1)\mho}\right),&
\end{align}
where $\mho$ is the mean signal power which is given as $\mho=\mathbb{E}[R_{\kappa\mathcal{F}}^{2}]=2\mu\sigma^{2}+d^{2}$, where $d^{2}=\sum_{i=1}^{\mu}p_{i}^{2}+q_{i}^{2}$. It is clear from \eqref{kf} and \eqref{akf} that $R_{\alpha\kappa\mathcal{F}}^{\alpha}=R_{\kappa\mathcal{F}}^{2}$ with mean
signal power  $\Omega=\mathbb{E}[R_{\alpha\kappa\mathcal{F}}^{2}]=\mathbb{E}[R_{\kappa\mathcal{F}}^{4\over\alpha}]$. Hence, the envelope PDF of the $\alpha$-$\kappa$-$\mathcal{F}$ composite distribution can be derived as
\begin{align}\label{akf1}
&f_{R_{\alpha\kappa\mathcal{F}}}(r) = {\alpha\mu^{\mu}(1+\kappa)^{\mu}\left((m_s-1)\omega\Omega^{\alpha\over2}\right)^{m_s}\exp{(-\mu\kappa)}\over {B}\left(\mu,m_s\right)\left(\mu (1+\kappa) r^{\alpha} +(m_s-1)\omega\Omega^{\alpha\over2} \right)^{\mu+m_s}}\cr&\times r^{\alpha\mu-1}
{}_{1}F_{1}\left(\mu+m_s;\mu;\frac{\mu^{2} \kappa(1+\kappa) r^{\alpha}}{\mu(1+\kappa)r^{\alpha} +(m_s-1)\omega\Omega^{\alpha\over2}}\right).&
\end{align}
With the help of $\gamma={\overline{\gamma}R_{\alpha\kappa\mathcal{F}}^{2}\over\mathbb{E}[R_{\alpha\kappa\mathcal{F}}^{2}]}={\overline{\gamma}R_{\alpha\kappa\mathcal{F}}^{2}\over\Omega}$, the PDF in \eqref{pdfgamk} is obtained. Thus, the proof is completed.
\subsubsection*{A.5) Proof of Lemma 5}
Using the series representation for ${}_{1}F_{1}(\cdot)$ \cite[Eq. (9.14.1)]{i:ryz} and then applying \cite[Eq. (3.194.5)]{i:ryz}, we obtain \eqref{cpdfgamk}. Thus, the proof is completed. Due to space limitations, the proof of \eqref{kamp} and \eqref{hum} is omitted.
\end{appendices}
\bibliographystyle{IEEEtran}
\bibliography{prodff}

\begin{thebibliography}{10}
\providecommand{\url}[1]{#1}
\csname url@samestyle\endcsname
\providecommand{\newblock}{\relax}
\providecommand{\bibinfo}[2]{#2}
\providecommand{\BIBentrySTDinterwordspacing}{\spaceskip=0pt\relax}
\providecommand{\BIBentryALTinterwordstretchfactor}{4}
\providecommand{\BIBentryALTinterwordspacing}{\spaceskip=\fontdimen2\font plus
\BIBentryALTinterwordstretchfactor\fontdimen3\font minus
  \fontdimen4\font\relax}
\providecommand{\BIBforeignlanguage}[2]{{%
\expandafter\ifx\csname l@#1\endcsname\relax
\typeout{** WARNING: IEEEtran.bst: No hyphenation pattern has been}%
\typeout{** loaded for the language `#1'. Using the pattern for}%
\typeout{** the default language instead.}%
\else
\language=\csname l@#1\endcsname
\fi
#2}}
\providecommand{\BIBdecl}{\relax}
\BIBdecl

\bibitem{8166770}
S.~K. {Yoo}, N.~{Bhargav}, S.~L. {Cotton}, P.~C. {Sofotasios}, M.~{Matthaiou},
  M.~{Valkama}, and G.~K. {Karagiannidis}, ``The $\kappa$-$\mu$/inverse gamma
  and $\eta$-$\mu$/inverse gamma composite fading models: Fundamental
  statistics and empirical validation,'' \emph{IEEE Trans. Commun.}, no. Dec.,
  pp. 1--16, 2017, early access article.

\bibitem{m:yac}
M.~D. Yacoub, ``The $\kappa$-$\mu$ distribution and the $\eta$-$\mu$
  distribution,'' \emph{{IEEE} Antennas Propag. Mag.}, vol.~49, no.~1, pp.
  68--81, Feb. 2007.

\bibitem{7460203}
H.~{Al-Hmood} and H.~S. {Al-Raweshidy}, ``Unified modeling of composite
  $\kappa$ -$\mu$ /gamma, $\eta$ -$\mu$/gamma, and $\alpha$ -$\mu$/gamma fading
  channels using a mixture gamma distribution with applications to energy
  detection,'' \emph{IEEE Antennas Wirel. Propag. Lett.}, vol.~16, pp.
  104--108, Apr. 2017.

\bibitem{Paris}
J.~F. Paris, ``Statistical characterization of $\kappa$-$\mu$ shadowed
  fading,'' \emph{IEEE Trans. Veh. Technol.}, vol.~63, no.~2, pp. 518--526,
  Feb. 2014.

\bibitem{1905-00069}
P.~Ram{\'{\i}}rez{-}Espinosa and F.~J. L{\'{o}}pez{-}Mart{\'{\i}}nez, ``On the
  utility of the inverse gamma distribution in modeling composite fading
  channels,'' in \emph{IEEE GlobeCom}, Waikoloa, HI, USA, Dec. 2019, pp. 1--6.

\bibitem{1904-05587}
P.~Ram{\'{\i}}rez{-}Espinosa, J.~M.~M. Moualeu, D.~B. da~Costa, and F.~J.
  L{\'{o}}pez{-}Mart{\'{\i}}nez, ``The $\alpha$-$\kappa$-$\mu$ shadowed fading
  distribution: Statistical characterization and applications,'' in \emph{IEEE
  GlobeCom}, Waikoloa, HI, USA, Dec. 2019, pp. 1--6.

\bibitem{Fisher}
S.~K. Yoo, S.~Cotton, P.~Sofotasios, M.~Matthaiou, M.~Valkama, and
  G.~Karagiannidis, ``The {F}isher-{S}nedecor $\mathcal{F}$ distribution: A
  simple and accurate composite fading model,'' \emph{IEEE Commun. Lett.},
  vol.~21, no.~7, pp. 1661--1664, Jul. 2017.

\bibitem{8920091}
K.~{Peppas}, G.~{Alexandropoulos}, E.~D. {Xenos}, and A.~{Maras}, ``The
  {Fischer-Snedecor} $\mathcal{F}$-distribution model for turbulence-induced
  fading in free-space optical systems,'' \emph{J. Lightw. Technol.}, vol.~38,
  no.~6, pp. 1286--1295, Mar. 2020.

\bibitem{4100514}
G.~{Fraidenraich} and M.~D. {Yacoub}, ``The $\alpha$-$\eta$-$\mu$ and
  $\alpha$-$\kappa$-$\mu$ fading distributions,'' in \emph{Proc. IEEE ISSSTA},
  Aug. 2006, pp. 16--20.

\bibitem{i:ryz}
I.~S. Gradshteyn and I.~M. Ryzhik, \emph{Table of Integrals, Series, and
  Products}, 7th~ed.\hskip 1em plus 0.5em minus 0.4em\relax Academic Press,
  California, 2007.

\bibitem{Srivastava}
H.~M. Srivastava and P.~W. Karlsson, \emph{Multiple Gaussian Hypergeometric
  Series}, 1st~ed.\hskip 1em plus 0.5em minus 0.4em\relax Hoboken, NJ, USA:
  Wiley, 1985.

\end{thebibliography}
\end{document}